\documentclass[12pt]{article}
\usepackage{amsmath}
\usepackage{amssymb}
\begin{document}
\vskip 2cm
\begin{center}
{\sf {\Large   Supervariable Approach to the Nilpotent Symmetries\\ for a
 Toy Model of the Hodge Theory}}

\vskip 3.0cm

{\sf D. Shukla$^{(a)}$, T. Bhanja$^{(a)}$, R. P. Malik$^{(a,b)}$}\\
$^{(a)}$ {\it Physics Department, Centre of Advanced Studies,}\\
{\it Banaras Hindu University, Varanasi - 221 005, (U.P.), India}\\

\vskip 0.1cm

$^{(b)}$ {\it DST Centre for Interdisciplinary Mathematical Sciences,}\\
{\it Faculty of Science, Banaras Hindu University, Varanasi - 221 005, India}\\
{\small {\sf {e-mails: dheerajkumarshukla@gmail.com; tapobroto.bhanja@gmail.com; rpmalik1995@gmail.com}}}

\end{center}

\vskip 2cm
\noindent
{\bf Abstract:} We exploit the standard techniques of the supervariable approach to derive
the nilpotent  Becchi-Rouet-Stora-Tyutin (BRST) and anti-BRST symmetry transformations for
a toy model of the Hodge theory (i.e. a rigid rotor) and  provide the geometrical meaning and 
interpretation to them. Furthermore, we {\it also} derive the nilpotent (anti-)co-BRST symmetry
transformations for this theory within the framework of the above supervariable approach.
We capture the (anti-)BRST and (anti-)co-BRST invariance of the Lagrangian of our present theory 
within the framework of augmented supervariable formalism. 
We also express the (anti-)BRST and (anti-)co-BRST charges in terms of the supervariables
(obtained after the application of the (dual-)horizontality conditions and (anti-)BRST and
(anti-)co-BRST invariant restrictions) to provide the geometrical interpretations for 
their nilpotency and anticommutativity properties. The application of the dual-horizontality 
condition and ensuing proper (i.e. nilpotent and absolutely anticommuting) fermionic
(anti-)co-BRST symmetries are completely {\it novel} results in our present investigation.
\vskip 2cm



\newpage
\section{Introduction}

The model of a rigid rotor has played a very decisive role in unraveling some of the deepest
mysteries of nature (especially in the context of atomic, molecular and nuclear physics). 
This model has also been
shown to be a prototype example of a gauge theory because it is endowed with the first-class
constraints in the language of Dirac's prescription for classification scheme [1,2]. As a 
consequence, it has also been discussed within the framework of Becchi-Rouet-Stora-Tyutin
(BRST) formalism for its quantization and constraint analysis (see, e.g. [3] for details).
We have shown, in our recent publication [4], that this toy model has a rich mathematical 
structure behind it because it provides a tractable physical example for the Hodge theory
(within the framework of the BRST formalism) where the continuous and discrete symmetries
of the theory provide the physical realizations of the de Rham cohomological operators
of differential geometry (see, e.g. [5-8]).

Two key mathematical properties, associated with the (anti-)BRST symmetries
(and their corresponding charges), are the nilpotency 
property and the absolute anticommutativity. The superfield approach to BRST formalism (see, e.g.[9-13]) 
provides the geometrical origin and interpretation for these abstract mathematical
properties in the language of the translational generators along the Grassmanian directions
of the supermanifold on which the ordinary gauge theories are generalized. This approach
has been applied in the context of a rigid rotor, too, so that the geometrical basis
for its (anti-)BRST symmetries could be provided (see, e.g. [4]). However, some 
unusual approximations have been made to derive the correct results. One of the purposes 
of our present investigation is to derive the nilpotent (anti-)BRST symmetry transformations in
a clear fashion on the basis of physically intuitive restrictions and provide 
the geometrical origin for them.

As has been pointed out earlier, the model of a rigid rotor is a physical example of Hodge 
theory within the framework of BRST formalism. Hence, there are nilpotent (anti-)BRST and 
(anti-)co-BRST symmetries in the theory (besides a unique bosonic and a ghost-scale symmetry).
 In this context, it is a challenging problem to provide a geometrical basis for the
 (anti-)co-BRST symmetry transformations within the framework of superfield approach to 
BRST formalism [5-8]. We resolve this issue in our 
present investigation by applying the augmented version of dual-horizontality condition (DHC)
and derive (not only the proper nilpotent (anti-)co-BRST symmetry transformations) but we also provide
the geometrical basis for their existence in the same manner as that of the (anti-)BRST
symmetries (which has already been done in our earlier work [4]). In the application of the
DHC, we exploit the working-rule, established in [14], for the Hodge duality $\star$ 
operation on a given supermanifold and obtain the precise results which establishes the correctness of 
the rules which have been laid down in our earlier publication [14].

In our present investigation, we have also provided the geometrical basis for the nilpotency
and absolute anticommutativity of the (anti-)co-BRST charges (on the same lines as we have
provided for the (anti-)BRST charges in our earlier work [4]). Furthermore, we also capture
the (anti-)BRST and (anti-)co-BRST invariance of the Lagrangian of our present theory within 
the framework of the augmented version of supervariable approach. This exercise leads to the
geometrical interpretation for the (anti-)BRST and (anti-)co-BRST invariance of the Lagrangian
in the language of the translation of a specific sum of composite supervariables
(obtained after the appropriate set of restrictions) along the Grassmannian directions 
of the chosen supermanifold on which our ordinary theory is generalized within the
 framework of the supervariable approach to BRST formalism.

Our present endeavor is essential on the following counts. First, in our earlier work [4], 
we have made some approximations to obtain the proper (anti-)BRST symmetry transformations 
within the framework of augmented supervariable approach. Thus, it is essential for us
 to derive the same symmetry transformations in a physically intuitive manner by exploiting
 the horizontality condition and (anti-)BRST invariant restrictions. We have accomplished 
this goal in our present endeavor. Second, to put the idea of the dual-horizontality condition
(DHC) on the firmer footings, it is necessary for us to apply it to our present system and derive the proper 
(anti-)co-BRST symmetry transformations. We have obtained these symmetry transformations in a consistent manner
by exploiting the idea of DHC. Finally, it is challenging for us to provide the geometrical basis for the
nilpotent (anti-)co-BRST transformations (and corresponding generators) within the framework of the supervariable
approach (as has already been done in [4] for the (anti-)BRST symmetries
 and their generators). We have achieved
this goal, too, in our present endeavor.

The material of our present paper is organized as follows. In Sec. 2, we briefly mention about
 the nilpotent (anti-)BRST and
(anti-)co-BRST symmetries for the Lagrangian of our present theory. Our Sec. 3 is devoted to the derivation
of nilpotent (anti-)BRST symmetries within the framework of augmented supervariable formalism. Sec. 4 
of our present endeavor contains the application of  dual-horizontality condition and 
the derivation of the (anti-)co-BRST symmetries.  Our Sec. 5 is devoted to capturing 
the geometrical meaning of the invariance of Lagrangian of our present theory under 
(anti-) BRST and (anti-)co-BRST transformations. In Sec. 6, we discuss the
 geometrical meaning of the nilpotency property of the (anti-)BRST and (anti-)co-BRST 
charges by expressing them in terms of the supervariables (obtained after various
appropriate restrictions). Finally, we make some concluding remarks and point out a
 few future directions in Sec. 7.

In our Appendix, we perform an explicit computation which is used in the main body of 
our text in the context of application of the dual-horizontality condition (DHC).

\noindent
\section{Preliminaries: Lagrangian and symmetries}

We begin with the following (anti-)BRST and (anti-)co-BRST invariant first-order 
Lagrangian for the rigid rotor  (see, e.g. [3,4] for details)
\begin{eqnarray}
L_b = \dot r \, p_r + \dot\vartheta \, p_{\vartheta} - \frac{p_{\vartheta}^2}{2r^2} -
 \lambda \,(r - a) + b\,(\dot \lambda - p_r)
+ \frac{b^2}{2} - i\, \dot{\bar C}\,\dot C + i\, \bar C\, C,
\end{eqnarray}
where $r$ and $\vartheta$ are the polar coordinates and their corresponding generalized velocities 
are $\dot r$ and $\dot\vartheta$. The momenta for the particle (of mass $m = 1$), moving on a circle of radius $a$,
are $p_r$ and $p_\vartheta$. Here $\lambda $ is a Lagrange multiplier that turns out to be the ``gauge'' variable
of our present theory. The variable $b$ is the Nakanishi-Lautrup type of auxiliary variable and $(\bar C)\,C$ 
are the (anti-)ghost fermionic $(C^2 = {\bar C}^2 = 0,\,\, C\,\bar C + \bar C\,C = 0)$ variables. All these variables 
are function of the evolution parameter {\it t} and an overdot on the variables always denotes the derivative
w.r.t. it (i.e. $ \dot{\vartheta} = d\vartheta/ dt, \,\dot{\lambda} = d\lambda/ dt $, etc).

We observe that under the following nilpotent  $(s_b ^2 = s_{ab}^2 = 0)$ and absolutely anticommuting
 $( s_{b}\,s_{ab} + s_{ab}\,s_{b} = 0 ) $ continuous (anti-)BRST symmetry transformations ($s_{(a)b}$):
\begin{eqnarray}
&& s_b\, \lambda = \dot C, \qquad \,\,s_b \, \bar C = i b, \quad\qquad \, s_b\, p_r = - C,\,\, 
\qquad s_b \,[C, r, \vartheta, p_{\vartheta}, b] = 0,\nonumber\\
&& s_{ab}\, \lambda = \dot {\bar C}, \qquad  s_{ab}\, C = - i b, \qquad  s_{ab}\, p_r = - \bar C,
\qquad  s_{ab}\, [\bar C, r, \vartheta, p_{\vartheta}, b] = 0,
\end{eqnarray}
the Lagrangian $L_b$ transforms to the total time derivatives:
\begin{eqnarray}
s_{ab}\, L_b = \frac{d}{dt}\bigl [b\,\dot {\bar C} - \bar C\,(r-a)\bigr],\qquad\quad
s_b\, L_b = \frac{d}{dt}\bigl [b\,\dot C - C\,(r-a)\bigr],
\end{eqnarray}
thereby rendering the action integral $S = \int dt L_b$ invariant. Hence, the transformations (2) 
are the {\it symmetry} transformations for the action $S$. There are other nilpotent  $(s_d ^2 = s_{ad}^2 = 0)$  
and absolutely anticommuting $ (s_{d}\,s_{ad} + s_{ad}\,s_{d} = 0) $ symmetries in the theory. 
These (anti-) co-BRST [or (anti-)dual-BRST] symmetry transformations ($s_{(a)d}$):
\begin{eqnarray}
&& s_d\, \lambda = \bar C, \qquad\,\,  s_d\, C = i\, (r-a), \,\quad\qquad  
s_d\, p_r = \dot{\bar C},\qquad\,\,  s_d \,[\bar C, r, \vartheta, p_{\vartheta}, b] = 0,\nonumber\\
&& s_{ad}\, \lambda = C, \qquad s_{ad}\, \bar C = - i\, (r - a), 
\qquad  s_{ad}\, p_r =  \dot C,\qquad  s_{ad}\, [ C, r, \vartheta, p_{\vartheta}, b] = 0,
\end{eqnarray}
leave the Lagrangian absolutely invariant (i.e. $s_{(a)d}\, L_b = 0$).

We have demonstrated that the action integral $S = \int dt L_b$ and Lagrangian ($L_b$) remain invariant under the 
continuous (anti-)BRST and (anti-)co-BRST symmetry transformations, respectively. Thus, according to Noether's theorem,
 the following conserved and nilpotent (anti-)BRST $(Q_{(a)b})$ and (anti-)co-BRST $(Q_{(a)d})$ charges, namely;
\begin{eqnarray}
&& Q_b = b\,\dot C - \dot b\, C, \qquad\quad Q_{ab} = b\,\dot {\bar C} - \dot b\,\bar C,\nonumber\\
&& Q_d = b\,\bar C + \dot b\, \dot{\bar C}, \qquad\quad Q_{ad} = b\, C + \dot b\,\dot C,
\end{eqnarray}
are the generators for the (anti-)BRST and (anti-)co-BRST symmetry transformations, as it can be explicitly checked that
\begin{eqnarray}
s_r \, \phi = \pm\, i\, [\phi, Q_r]_{\pm}, \qquad\qquad r = b, \,ab, \,d, \,ad,
\end{eqnarray}
for the generic variable $\phi = r, \vartheta, p_r, p_{\vartheta}, \lambda, b, C, \bar C$. 
Here $\pm$ signs, as the subscripts on the square bracket, correspond to the (anti)commutator for the 
generic variable $\phi$ of our theory being (fermionic) bosonic in nature.

We wrap up this section with the following remarks. First, under the (anti-)BRST symmetry transformations, 
it is the kinetic term $[ (\dot{\vartheta}\,p_{\vartheta}) - ({p^2_{\vartheta}}/{2r^2})] 
= \frac{1}{2}\,{\dot{r}}^2\,{\dot{\vartheta}}^2 = \frac{1}{2}\, v^2$ 
that remains invariant. Second, the gauge-fixing term $ (\dot{\lambda} - p_r) $ turns out to be invariant quantity 
under the nilpotent (anti-)co-BRST symmetry transformations. Third, the kinetic term $ \frac{1}{2}\, v^2 $ 
has its origin [4] in the exterior derivative $ d = dt\, \partial_t $ (with $ d^2 = 0 $). Fourth, 
the gauge-fixing term $ (\dot{\lambda} - p_r) $ owes its origin to the co-exterior derivative $ \delta = \ast\, d \,\ast $ 
(with $ \delta^2 = 0 $) of differential geometry [4] where ($\ast$) is the Hodge duality operation. 
Fifth, the anticommutator of the (anti-)BRST and (anti-)co-BRST symmetry transformations defines a unique bosonic  
symmetry in the theory which corresponds to the Laplacian operator of differential geometry. 
Finally, the present toy model of a rigid rotor turns out to be the physical example of a Hodge theory 
within the framework of BRST formalism [4].

\noindent
\section{(Anti-)BRST symmetries: Supervariable formalism}

In our earlier work [4], the (anti-)BRST symmetry transformations have been obtained by exploiting the basic
ideas of supervariable formalism. However, there have been ad-hoc assumptions and approximations in deriving
the correct results. In our present section, we exploit the horizontality condition and (anti-)BRST invariant 
restriction to obtain the appropriate (anti-)BRST symmetry transformations for our system without making any 
approximations. Our method of derivation is simpler and physically more intuitive. To corroborate these statements, 
first of all, we generalize the gauge and (anti-)ghost variables (i.e. $ \lambda(t), C(t), \bar{C}(t) $)
onto (1, 2)-dimensional supermanifold as supervariables:
\begin{eqnarray}
&& \lambda(t) \rightarrow {\Lambda}(t, \theta,\bar{\theta}) = \lambda (t) + \theta\, \bar{R}(t) + \bar{\theta}\, R(t) 
+ i\, \theta\, \bar{\theta}\, S(t),\nonumber\\
&& C(t) \rightarrow F(t, \theta, \bar{\theta}) = C(t) + i\,\theta \bar{B}_1 (t) + i\,\bar{\theta} \, B_1(t) 
+ i\,\theta\,\bar{\theta}\, s(t),\nonumber\\
&& \bar{C}(t) \rightarrow \bar{F}(t, \theta,\bar{\theta}) = \bar{C}(t) + i \theta\, \bar{B}_2(t) 
+ i\,\bar{\theta}\,B_2(t) + i\,\theta\,\bar{\theta} \,
\bar{s}(t).
\end{eqnarray}
where the expansions have been made along the Grassmannian directions 
$ (\theta, \bar{\theta}) $ of the (1, 2)-dimensional
supermanifold which is parametrized by the superspace variable $ Z^{M} = (t, \theta, \bar{\theta}) $ 
and the secondary variables ($ R, \bar{R}, s, \bar{s} $) are fermionic and
 ($ B_1, \bar{B}_1, B_2, \bar{B}_2,\,S $) are 
bosonic in nature. It is elementary to check that, in the limit $ \theta = 0 $, $ \bar{\theta} = 0 $,
we get back the original variables ($\lambda (t), C(t), \bar{C}(t)$) of our starting Lagrangian (1).
We christen the above supersymmetric generalized variables as ``supervariables'' (and {\it not}
{\it superfields}) because, in the limit $ \theta = \bar\theta = 0$, we retrieve back our basic  dynamical variables
(and {\it not} the fields).

In one (0 + 1)-dimensional ordinary space, we note that the 1-forms 
$d = dt \, \partial_t,\,\lambda^{(1)} = dt\, \lambda (t),$ 
lead to the definition of a 2-form $ d\, \lambda^{(1)} = (dt \wedge dt)\, \dot{\lambda} = 0$ 
where $ d = dt\,\partial_t$ is
the exterior derivative (with $ d^{2} = 0 $) and $ (dt \wedge dt) = 0 $. 
These operators can be generalized onto (1, 2)-dimensional supermanifold to their supersymmetric counterparts as
\begin{eqnarray}
&& d \longrightarrow \tilde{d} = dZ^{M}\,\partial_{M} \equiv dt\, \partial_t + d\theta \, \partial_\theta 
+ d \bar{\theta}\,\partial_{\bar{\theta}},\quad\qquad\quad {\tilde{d}}^{2} = 0, \nonumber\\
&& \lambda^{(1)} \longrightarrow {\tilde{\lambda}}^{(1)} = dZ^{M}\, A_{M} \equiv dt\,\Lambda (t, \theta, \bar{\theta}) 
+ d\theta \, \bar{F}(t, \theta, \bar{\theta}) + d\bar{\theta}\, F(t, \theta, \bar{\theta}),
\end{eqnarray}
where the supervariables 
$ (\Lambda(t, \theta, \bar{\theta}),\,F(t, \theta, \bar{\theta}),\,\bar{F}(t, \theta, \bar{\theta})) $ 
form a vector supermultiplet $ A_{M} $ on the (1, 2)-dimensional supermanifold whose expansions along the 
Grassmanian directions have been given in (7). In the above, we have taken 
$ \partial_{M} = \partial/\partial{Z^{M}}  \equiv ( \partial_{t},\, \partial_{\theta},\, \partial_{\bar{\theta}}) $ 
as the derivatives w.r.t. the evolution parameter {\it t} and the Grassmanian variables $ (\theta,\, \bar{\theta}) $. 
The super 2-form, constructed with $ \tilde{d} $ and $ \tilde{\lambda}^{(1)}$, has the following explicit form:
\begin{eqnarray}
&& \tilde{d}\, \tilde{\lambda}^{(1)} = (dt \wedge dt)\, (\partial_t \,\bar{F} - \partial_\theta\, \Lambda)
+ (dt \wedge d\bar{\theta})\, (\partial_t\, F - \partial_{\bar{\theta}}\, \Lambda) 
+ (d\theta \wedge d\theta) \, (\partial_{\theta}\, \bar{F}) \nonumber\\
&& \qquad \quad+ (d\bar{\theta}\wedge d\bar{\theta})\,(\partial_{\bar{\theta}}\, F) 
+ (d \theta \wedge d\bar{\theta})\, (\partial_{\theta}\, F + \partial_{\bar{\theta}}\, \bar{F}).
\end{eqnarray}
The horizontality condition requires that $ d\,\lambda^{(1)} = \tilde{d}\, {\tilde{\lambda}}^{(1)} = 0 $. 
Thus, we obtain the following expressions for the secondary variables in terms if the basic dynamical and 
auxiliary variables, namely; (see, e.g. [4] for details)
\begin{eqnarray}
&& R = \dot{C},\qquad \bar{R} = \dot{\bar{C}},\qquad S = \dot{b},\,\,\quad\qquad {\bar{B}}_{2} = 0,\nonumber\\
&& B_{1}= 0,\qquad s = 0,\qquad \,\, \bar{B}_{1} + B_{2} = 0, \quad \bar{s} = 0.
\end{eqnarray}
The condition $\bar{B}_{1} + B_{2} = 0  $ is nothing but the Curci-Ferrari type restriction which is trivial in
our case. Thus, we choose $ B_{2} = - {\bar{B}}_{1} = b $. This specific choice can 
be derived using the (anti-)BRST invariant restriction, too. 
As a consequence, we have the following expansions for
the supervariables after the application of the horizontality condition (HC):
\begin{eqnarray}
\Lambda^{(h)}(t, \theta, \bar{\theta}) &=& \lambda(t) + \theta\, (\dot{\bar{C}}) + \bar{\theta}\,(\dot{C}) + \theta\,\bar{\theta}\, 
(i \dot{b})\nonumber\\
&\equiv & \lambda (t) + \theta\,(s_{ab}\,\lambda) + \bar{\theta}\, (s_{b} \, \lambda) + \theta\,\bar{\theta}\, (s_{b}\,s_{ab}\, \lambda), \nonumber\\
F^{(h)}(t, \theta, \bar{\theta}) &=& C(t) + \theta\, (- i b) + \bar{\theta}\,(0) + \theta\bar{\theta}\, (0) \nonumber\\
&\equiv & C(t) + \theta\, (s_{ab\, C}) + \bar{\theta}\, (s_{b}\,C) + \theta\,\bar{\theta}\, (s_{b}\, s_{ab}\, C),\nonumber\\
{\bar{F}}^{(h)}(t, \theta, \bar{\theta}) &=& \bar{C}(t) + \theta\, (0) + \bar{\theta}\, (i b) + \theta\bar{\theta}\, (0)\nonumber\\
&\equiv & \bar{C}(t) + \theta \,(s_{ab}\,\bar{C}) + \bar{\theta}\, (s_{b}\,\bar{C}) + \theta\,\bar{\theta}\, (s_{b} \,s_{ab}\, \bar{C}).
\end{eqnarray}
where the superscript $(h)$, on the supervariables, denotes the super-expansions, obtained after the application 
of the HC. It is evident, from the above, that we have already derived the (anti-)BRST symmetry transformations 
for the variables $ (\lambda(t), C(t), \bar{C}(t)) $.

To derive the (anti-)BRST symmetries for the momentum variable $ p_{r}(t) $, we have to exploit the (anti-)BRST
invariant restrictions (BIRs). In this context, we note that the following (anti-)BRST invariant quantity (i.e.
quantity present in the square bracket)
\begin{equation}
s_{(a)b} \left[ \,b (t)\, p_{r}(t) - i\, \bar{C}(t)\,C(t)\right] = 0,
\end{equation}
can be generalized onto the (1, 2)-dimensional supermanifold as:
\begin{equation}
\,B(t, \theta, \bar{\theta})\, P_{r}(t, \theta, \bar{\theta}) 
- i \,{\bar{F}}^{(h)}(t, \theta, \bar{\theta})\,F^{(h)}(t, \theta, \bar{\theta}).
\end{equation}
The (anti-)BRST invariance of the Nakanishi-Lautrup auxiliary variable $ b(t) $ [i.e. $ s_{(a)b}\, b (t) = 0 $] 
implies that $ b(t) \to B(t, \theta, \bar{\theta})= b (t)  $. In other words, the supervariable 
$ B(t, \theta, \bar{\theta}) $ 
would have {\it no} expansion along the Grassmanian directions ($ \theta, \bar{\theta} $). 
To proceed further, we take the general 
expansion for the supervariable $ P_{r}(t, \theta, \, \bar{\theta})$ as:
\begin{equation}
P_{r}(t, \theta, \bar{\theta}) = p_{r}(t) + \theta \,\bar{K}(t) + \bar{\theta}\, K(t)
 + i\, \theta\,\bar{\theta}\, L(t),
\end{equation}
and demand that the (anti-)BRST invariant quantity $ (b \, p_{r} - i\, \bar{C}\,C) $ 
should  remain {\it independent} of the``soul" coordinates $ (\theta,\, \bar{\theta}) $. 
It will be noted that in the old literature on superfield 
approach to BRST formalism [13], the bosonic coordinates have been christened as the ``body''
coordinates and Grassmannian coordinates have been called as the ``soul'' coordinates because the latter 
are very abstract and can not
be physically realized in the ordinary space. 
In other words, we impose the following restriction:
\begin{equation}
b (t)\, P_{r}(t, \theta,\, \bar{\theta}) - i\, {\bar{F}}^{(h)}(t, \theta, \bar{\theta})
\,F^{(h)}(t, \theta,\, \bar{\theta}) =  b (t)\,p_{r}(t) - i \,\bar{C}(t)\,C(t),
\end{equation}
which leads to the determination of the secondary variables of (14) in terms of the basic and auxiliary variables 
of our present theory as:
\begin{equation}
\bar{K} = - \,\bar{C},\qquad \quad K = -\,C, \qquad \quad L = -\,b.
\end{equation}
The above results establish the bosonic nature of $ L $ and fermionic nature of ($ K, \bar{K} $). 
Thus, we have the following expansions for $ P_{r}(t, \theta, \bar{\theta}) $:
\begin{eqnarray}
P^{(b)}_r(t, \theta, \bar{\theta}) &=& p_{r}(t) + \theta\, (-\bar{C}) + \bar{\theta}\, (-C) 
+ \theta\bar{\theta}\, (-i b)\nonumber\\
&\equiv & p_{r}(t) + \theta\,(s_{ab}\,p_{r}) + \bar{\theta}\,(s_{b}\,p_{r}) 
+ \theta\bar{\theta}\, (s_{b}\,s_{ab}\, p_{r}),
\end{eqnarray}
where the superscript $ (b) $ stands for the supervariable obtained after the application of the (anti-)BRST 
invariant restriction (15). It is evident, we have found out the (anti-)BRST symmetry 
transformations for $ p_{r}(t) $, as:
\begin{equation}
s_{b}\, p_{r} = -\, C,\quad\qquad s_{ab}\, p_{r} = -\,\bar{C},\quad\qquad s_{b}\,s_{ab}\, p_{r} = -i\,b.
\end{equation}
We conclude that, for the derivation of the correct and complete set of (anti-)BRST symmetries, 
we have to exploit the HC and BIR 
(cf. (15)) {\it together}.

We close this section with the remarks that we have obtained the proper (anti-)BRST symmetry transformations
for {\it all} the variables $ (\lambda, C, \bar{C}, p_{r}) $ of our theory. Our method of derivation of these 
(anti-)BRST symmetries is more {\it physical} in content than the same derivation carried out in our earlier 
work [4]. The key ideas that have been exploited {\it together} in our present endeavor are the HC and 
(anti-)BRST invariant restrictions (BIRs) which lead to the derivation of the {\it full} 
set of proper (anti-)BRST symmetries. 
A close look at the expansions (11) and (17) demonstrate that
\begin{eqnarray}
\frac{\partial}{\partial\theta}\,\Omega^{(h,b)}(t, \theta, \bar{\theta})\mid_{\bar{\theta} = 0} \,=
\, s_{ab}\,\omega(t),\quad\qquad \frac{\partial}{\partial\bar{\theta}}\,
\Omega^{(h,b)}(t, \theta, \bar{\theta})\mid_{\theta = 0}\, = \, s_{b}\,\omega(t),
\end{eqnarray}
where $ \omega(t) $ is the ordinary one (0 + 1)-dimensional variable and $ \Omega^{(h,b)}(t, \theta, \bar{\theta}) $ 
are the supervariables obtained after HC and BIRs (cf. (11), (17)). 
The above relationships provide the geometrical meaning for the (anti-)BRST symmetry transformations $ s_{(a)b} $. 
It states that the (anti-)BRST symmetry transformations $s_{(a)b}$.  
of an ordinary variable $ \omega (t) $ is equivalent
 to the translations of the corresponding supervariable (cf. (11), (17)) along the Grassmanian directions 
$(\theta,\,\bar\theta)$ of the (1, 2)-dimensional supermanifold. The nilpotency of 
$s_{(a)b}$ (i.e. \,${s^{2}_{(a)b}} = 0)$
is connected with two successive translations (i.e. $\partial^{2}_\theta = 0,\,\partial^{2}_{\bar\theta} = 0$) 
along the Grassmanian directions $(\theta,\,\bar\theta)$ of our chosen (1, 2)-dimensional supermanifold.

\noindent
\section{Nilpotent (anti-)co-BRST symmetries: Supervariable approach}

In our present section, we shall exploit the concept of dual-horizontality condition to derive
the nilpotent (anti-)co-BRST symmetry transformations (cf. Sec. 2). The latter are
characterized by the key observation that the gauge-fixing term $(\dot{\lambda} - p_r )$ remains invariant
under it. Thus, it is clear from the key ideas of the augmented version of the supervariable
approach that this quantity would remain independent of the ``soul'' coordinates $(\theta,\, \bar{\theta})$
when it is generalized onto the (1, 2)-dimensional supermanifold. Towards this goal in mind,
first of all, we note that the following is true, namely;
\begin{equation}
\delta{\lambda}^{(1)} = \ast\, d\,\ast \,{\lambda}^{(1)} = \dot{\lambda},
\end{equation}
where $ \ast\, d\,\ast $ is the co-exterior derivative, ${\lambda}^{(1)} = dt\,\lambda(t)$ is the 1-form in one 
(0 + 1)-dimensional ordinary space and $(\ast)$ is the Hodge-duality operation on 1D ordinary spacetime manifold. 
The invariance of the gauge-fixing term under the (anti-)co-BRST symmetry transformations can be translated into 
the following (anti-)co-BRST invariant restriction (CBIR) on the supervariables of the (1, 2)-dimensional supermanifold:
\begin{equation}
\star\, \tilde{d}\,\star\, {\tilde{\lambda}}^{(1)} - P_{r}(t, \theta, \bar{\theta}) = \ast\, d\,\ast \,{\lambda}^{(1)} - p_{r}(t),
\end{equation}
where ($ \star $) is the Hodge-duality operation on the (1, 2)-dimensional supermanifold and
$ \tilde{\lambda}^{(1)},\, \tilde{d},\,P_{r}(t, \theta, \bar{\theta}) $ are defined in equations (8) and (14). 
We christen the CBIR (21) as the dual-horizontality condition (DHC) because it is the co-exterior derivative of 
differential geometry that plays a key role in the above restriction.

We have the step-by-step computation of $ \star\, \tilde{d}\,\star\, {\tilde{\lambda}}^{(1)} $ in our Appendix A. 
Ultimately, the DHC (cf. (21)) leads to the following equality:
\begin{eqnarray}
(\dot{\Lambda} + \partial_{\theta}\,\bar{F} + \partial_{\bar{\theta}}\, F) &+& s^{\theta\,\theta} \, (\partial_{\theta}\,F) 
+  s^{\bar\theta\,\bar\theta}\,(\partial_{\bar{\theta}}\,\bar{F}) \nonumber\\
&-& [p_{r}(t) + \theta\, \bar{K}(t) + \bar{\theta}\, K(t) + i\,\theta\,\bar{\theta}\, L(t)] = \dot{\lambda} - p_{r}(t).
\end{eqnarray}
It is clear that the coefficients of $ s^{\theta\,\theta}$ and $ s^{\bar\theta\,\bar\theta} $ would be zero because there 
are  no such terms
on the r.h.s. Thus, we have the following results, namely;
\begin{eqnarray}
\partial_{\theta}\,F = 0,\quad \partial_{\bar\theta}\,\bar F = 0, \Longrightarrow \bar B_{1} = 0, \quad B_2 = 0, \quad s = 0,\quad \bar s = 0. 
\end{eqnarray}
The above values imply that the reduced form of the expansions for 
$F(t, \theta, \bar{\theta}) $ and $\bar F(t, \theta, \bar{\theta})$ (cf. (7)) are as given below:
\begin{eqnarray}
F^{(r)}(t, \theta, \bar{\theta}) = C(t) + i\,\bar\theta\, B_1(t), \qquad  \bar{F}^{(r)}(t, \theta, \bar{\theta}) = \bar C(t) + i\,\theta\, \bar B_2(t).
\end{eqnarray}
Plugging in these expansions and that of $\Lambda(t, \theta, \bar{\theta}) $ from (7) into the CBIR (cf. (21)), 
we obtain the following conditions on the secondary variables:
\begin{eqnarray} 
B_1 +  \bar B_2 = 0, \qquad \bar K = \dot{\bar R}, \qquad K = \dot R, \qquad L = \dot S.
\end{eqnarray}
In the above, the condition $B_1 + \bar B_2 = 0$ is the analogue of Curci-Ferrari restriction. Making the choice
$B_1(t) = {\cal B}$, we get $\bar B_2(t) = -\,{\cal B}$. Substitution of these values into expansions (7) and (24)
lead to the following re-reduced form of these expansions:
\begin{eqnarray} 
&& F^{(R)}(t, \theta, \bar{\theta}) = C(t) + i\,\bar\theta\, {\cal B},\nonumber\\
&& {\bar {F}}^{(R)}(t, \theta, \bar{\theta}) = \bar C(t) - i\,\theta\, {\cal B},\nonumber\\
&& P^{(R)} (t, \theta, \bar{\theta}) = p_{r}(t) + \theta \,(\dot{\bar R}) + \bar\theta\, (\dot R) 
+ i\,\theta\,\bar\theta (\dot S). 
\end{eqnarray}
The above expansions show that we have {\it not yet} found the explicit expressions for the
secondary variables in terms of the basic and auxiliary variables.

The additional restrictions come from the following observations, namely;
\begin{eqnarray} 
s_{(a)d}\,\bigl[ \dot r\,p_r - i\, \dot{\bar C}\, \dot C\bigr] = 0, \qquad\qquad s_{(a)d}\,\bigl[ \lambda\,(r - a)  
- i\, \bar{C}\, C\bigr] = 0.
\end{eqnarray}
The above (anti-)dual BRST invariant quantities (which are present in the square brackets) can be generalized onto 
(1, 2)-dimensional supermanifold. By exploiting the idea of augmented version of supervariable approach, we have to 
demand that such invariant quantities should be independent of the Grassmannian 
variables $ \theta $ and $ \bar{\theta} $.
 Thus, we have the following equality
 conditions on the supervariables of our chosen supermanifold:
\begin{eqnarray}
 &&\dot{R} (t, \theta, \bar\theta) \, P^{(R)}_{r} (t, \theta, \bar\theta)
- i\,\dot{\bar{F}}^{(R)} (t, \theta, \bar\theta)\, 
\dot{F}^{(R)} (t, \theta, \bar\theta) 
= \dot{r}\,p_{r} - i\,\dot{\bar{C}}\,\dot{C},\nonumber\\
&&\Lambda (t, \theta, \bar\theta) \,[R (t, \theta, \bar\theta) - a] 
- i\,\bar{F}^{(R)} (t, \theta, \bar\theta) \, F^{(R)} (t, \theta, \bar\theta)  =
\,\lambda \,(r - a) - i\, \bar{C}\,C,
\end{eqnarray}
where the expressions for $ (P^{(r)}_{r},\, F^{(R)},\,\bar{F}^{(R)}) $ are given in (26) and $ R(t, \theta, \bar{\theta}) $ 
is the generalization of $ r(t) $ onto (1, 2)-dimensional supermanifold. However, as we  know that $ r(t) $ is 
an (anti-)co-BRST invariant (i.e. $ s_{(a)d}\,r(t) = 0  $) quantity, we find that $R(t, \theta, \bar{\theta}) = r(t)  $. 
Plugging in these values into (28) and $\Lambda (t, \theta, \bar \theta)$  from (7), the above equality becomes:
\begin{eqnarray}
 && \dot{r}\,\left[ p_{r} + \theta\,(\dot{\bar{R}}) + \bar{\theta}\,(\dot{R}) + i \theta\,\bar{\theta}\,(\dot{S})\right] 
  - i \,\left[ \dot{\bar{C}} 
 - i\,\theta\,\dot{\cal B}\right] \,\left[\dot{C} + i\,\bar{\theta}\,\dot{\cal B}\right] 
  = \dot{r}\,p_{r} - i\, \dot{\bar{C}}\,\dot{C},\nonumber\\
 && \left[ \lambda + \theta\, \bar{R} + \bar{\theta}\, R
 + i\,\theta\,\bar{\theta}\, S\right]\,(r - a) - i \,\left[ {\bar{C}} 
 - i\,\theta\,{\cal B}\right] \,\left[{C} + i\,\bar{\theta}\,{\cal B}\right] 
 =  \,\lambda\,(r - a) - i\, \bar{C}\, C.
\end{eqnarray}
The above two equations in (29) yield the following beautiful relationships:
\begin{eqnarray}
&& \dot{\cal B}\, \dot{C} = \dot{\bar{R}}\, \dot{r},\qquad\qquad\,\,\,\, \dot{\cal B}\,\dot{\bar{C}} 
= \dot{R}\, \dot{r}, \qquad\qquad \,\,\,\,\dot{\cal B}\,\dot{\cal B} = \, \dot{S}\,\dot{r},\nonumber\\
&& {\cal B}\, C = \bar{R}\,(r - a),\qquad {\cal B}\,\bar{C} = R \, (r - a),\qquad \,\, {\cal B}\,{\cal B} = \, S\, (r - a).
\end{eqnarray}
Even after the relations in (30), we have {\it not} found the precise expressions for the secondary
variables in terms of the basic and auxiliary variables of the theory. Thus, we have to look for other (anti-)co-BRST
invariant quantities of our present theory.

We note that $s_d \,[\lambda \,\bar C] = 0 $ and $s_{ad}\, [\lambda\,  C] = 0 $. These co-BRST and anti-co-BRST
invariant quantities would also be independent of the ``soul'' coordinates $\theta$ and  $\bar\theta$ when
they are generalized onto the (1, 2)-dimensional supermanifold. Thus, we have the following restrictions on the 
supervariables of the above supermanifold, namely;
\begin{eqnarray}
\Lambda\, {\bar F}^{(R)} = \lambda\, \bar C, \qquad\qquad \Lambda\, {F}^{(R)} = \lambda\, C.
\end{eqnarray}
Plugging in the expansions from (7) and (26), we obtain the following relationships:
\begin{eqnarray}
&& R\,\bar C = 0, \quad\qquad R\,{\cal B} = S\, \bar C, \quad\qquad \bar R\,\bar C = i\,\lambda\,{\cal B},\nonumber\\
&& \bar R\, C = 0,\quad \qquad \bar R\,{\cal B} = S\, C,\quad \qquad R\,C = -\, i\,\lambda\,{\cal  B}.
\end{eqnarray}
The above relationships fix the value of $R$ and $\bar R$ as $R \propto \bar C $ and $\bar R \propto C$. 
We make, one of the simplest choices for the secondary variables as: $R = \bar C $ and  $\bar R = C$. Once we make this simple choice, the
rest of the secondary variables of the super-expansion also get fixed.

A careful observation of the above relationships (30) and (32), lead to the following expressions for
the secondary variables in terms of basic variables of our theory, namely;
\begin{equation}
\bar{R} = C,\quad\qquad R = \bar{C},\qquad\quad {\cal B} = (r - a),\qquad\quad S = (r - a)\,\equiv {\cal B}.
\end{equation}
Substitution of these values into expansions (7) and (26), lead to the following:
\begin{eqnarray}
\Lambda^{(d)}\,(t, \theta, \bar{\theta}) &=& \lambda(t) + \theta\, (C) + \bar{\theta}\, (\bar{C})  +  \theta\, \bar{\theta}\,
[i\, (r - a)] \nonumber\\
&\equiv & \lambda(t) + \theta\, (s_{ad}\, \lambda) + \bar{\theta}\,(s_{d}\, \lambda) + \theta\,\bar{\theta}\, (s_{d}\,s_{ad}\,\lambda), \nonumber\\
F^{(d)}\,(t, \theta, \bar{\theta}) &=& C(t) + \theta\, (0) + \bar{\theta}\, [i\,(r - a)]  +  \theta\, \bar{\theta}\, (0) \nonumber\\
&\equiv & C(t) + \theta\, (s_{ad}\, C) + \bar{\theta}\,(s_{d}\, C) + \theta\,\bar{\theta}\, (s_{d}\,s_{ad}\,C), \nonumber\\
{\bar F}^{(d)}\,(t, \theta, \bar{\theta}) &=& \bar{C}(t) + \theta\, [-\,i\,(r - a)] + \bar{\theta}\,(0)  +  \theta\, \bar{\theta}\, (0) \nonumber\\
&\equiv & \bar{C}(t) + \theta\, (s_{ad}\, \bar{C}) + \bar{\theta}\,(s_{d}\, \bar{C}) + \theta\,\bar{\theta}\, (s_{d}\,s_{ad}\,\bar{C}),\nonumber\\
P^{(d)}_{r}(t, \theta, \bar{\theta}) &=& p_{r}(t) + \theta\, (\dot{C}) + \bar{\theta}\, (\dot{\bar{C}}) 
+ \theta\,\bar{\theta}\, (i \dot r)\nonumber\\
& \equiv & p_{r}(t) + \theta\,(s_{ad}\,p_{r}) + \bar{\theta}\, (s_{d}\,p_{r}) + \theta\,\bar{\theta}\,(s_{d}\,s_{ad}\,p_{r}),
\end{eqnarray}
where the superscript $(d)$ denotes the expansion of the supervariables after the application of DHC.
We point out that we have already derived the 
nilpotent and absolutely anticommuting (anti-)co-BRST symmetry transformations (4)
in the above super-expansions. 
We note, from the above expressions, that there is a deep connection between the (anti-)co-BRST symmetry
transformations $s_{(a)d}$ and the translational generators ($ \partial_{\theta},\,\partial_{\bar\theta}$)
along the Grassmannian directions of the (1, 2)-dimensional supermanifold on which our theory has been 
generalized. In fact, we have the following mappings:
\begin{eqnarray}
&&\frac{\partial}{\partial\theta}\, \Sigma^{(d)}(t,\,\theta,\,\bar\theta)|_{\bar\theta = 0} = s_{ad}\, \sigma(t),
\qquad \qquad \quad \frac{\partial}{\partial\theta} \,\,\Longleftrightarrow \,\, s_{ad}, \nonumber\\
&& \frac{\partial}{\partial{\bar\theta}}\, \Sigma^{(d)}(t,\,\theta,\,\bar\theta)|_{\theta = 0} = s_{d}\, \sigma(t),
\qquad \qquad \quad \;\frac{\partial}{\partial\bar \theta}\,\, \Longleftrightarrow \,\, s_{d}, \nonumber\\
&&\frac{\partial}{\partial{\bar\theta}}\,\frac{\partial}{\partial\theta}\, \Sigma^{(d)}(t,\,\theta,\,\bar\theta) = s_{d}\,s_{ad} \,\sigma(t), \qquad \qquad \frac{\partial}{\partial\bar \theta} \frac{\partial}{\partial\theta}
 \,\,\Longleftrightarrow \,\, s_d s_{ad},
\end{eqnarray}
where $\sigma(t)$ is the generic variable of 1D ordinary space and $ \Sigma^{(d)}(t,\,\theta,\,\bar\theta)$ is 
the generic supervariable that is obtained in (34) with full super-expansions.

Geometrically, we note that the co-BRST symmetry transformations on a
given variable $\sigma(t)$ of the 1D theory is equivalent to the translation of the corresponding supervariable
$ \Sigma^{(d)}(t,\,\theta,\,\bar\theta)$ along the $\bar\theta$ direction of the supermanifold (where the 
Grassmannian direction $\theta$ is kept untouched).  Similarly, the geometrical origin and interpretation for the
anti-co-BRST symmetry transformation can be provided. We further lay emphasis on the observation that the 
nilpotency ($s_{(a)d}^{2} = 0$) and absolute anticommutativity ($s_d\,s_{ad} + s_{ad}\,s_{d} = 0$) properties 
of the (anti-)co-BRST symmetry transformations $s_{(a)d}$ are deeply connected with such properties 
(i.e. ${\partial_\theta}^2 = {\partial_{\bar\theta}}^2 = 0, \, \partial_\theta\,\partial_{\bar\theta}
+ \partial_{\bar\theta}\,\partial_{\theta} = 0$) associated with the translational
generators $\partial_\theta$ and $\partial_{\bar\theta}$ on the
(1, 2)-dimensional supermanifold on which our present theory is generalized. Thus, it is the supervariable approach to 
BRST formalism that provides geometrical meaning to the abstract mathematical properties 
(e.g. nilpotency and absolute anticommutatvity) associated with the (anti-)BRST
symmetry transformations (and corresponding (anti-)BRST charges). Furthermore, this formalism
also provides the inter-relationships between nilpotency and anticommutativity
properties (as we shall see in Sec. 6 of our present endeavor).

\section{Invariance of Lagrangian: Supervariable approach}

First of all, we note that the starting Lagrangian (1) can be written in the following
{\it three} different and distinct forms:
\begin{eqnarray} 
L_b &=& \dot r\,p_r + \dot\vartheta\,p_{\vartheta} - \frac{{p^2_{\vartheta}}}{2r^2} -
 \lambda\,(r - a) +s_b\,\Bigl[-\,i\,\bar C\,\big\{(\dot\lambda- p_r) + \frac{b}{2}\bigr\}\Bigr],\nonumber\\
&\equiv &  \dot r\,p_r + \dot\vartheta\,p_{\vartheta} - \frac{{p^2_{\vartheta}}}{2r^2} -
 \lambda\,(r - a) +s_{ab}\,\Bigl[+\,i\, C\,\big\{(\dot\lambda- p_r) + \frac{b}{2}\bigr\}\Bigr],\nonumber\\
&\equiv &  \dot r\,p_r + \dot\vartheta\, p_{\vartheta} - \frac{{p^2_{\vartheta}}}{2r^2} -
 \lambda \,(r - a) + s_b\,s_{ab}\,\Bigl[\frac{i}{2}\,({\lambda}^2 - {p_r}^2) + \frac{C\, \bar C}{2}\Bigr],
\end{eqnarray}
where it is basically the gauge-fixing and Faddeev-Popov ghost terms that have been ex-
pressed in {\it three} different ways because the original Lagrangian for the rigid rotor
(without the gauge-fixing and Faddeev-Popov ghost terms) is (see, e.g. [3] for details):
\begin{eqnarray}
L_0 = \dot r\,p_r + \dot\vartheta\,p_{\vartheta} - \frac{{p_{\vartheta}}^2}{2r^2} - \lambda\,(r - a).
\end{eqnarray}
To be precise, all the above three forms are inter-connected because the 
top two forms can be obtained from the bottom relation if we exploit the absolute anticommutativity property
($s_b\,s_{ab} + s_{ab}\,s_{b} = 0$) of the nilpotent ($s_{(a)b}^2 = 0$)
(anti-)BRST symmetry transformations $s_{(a)b}$.
Towards our main goal of expressing the Lagrangian (1) in terms of the supervariables, 
obtained after the application of suitable restrictions, first of all, we generalize the original Lagrangian
$L_0$ onto (1, 2)-dimensional supermanifold as:
\begin{eqnarray}
L_0 \longrightarrow \tilde L_0 = \dot r\, {P_r}^{(b)}(t,\,\theta,\,\bar \theta) + \dot\vartheta\,p_{\vartheta}
- \frac{{p_{\vartheta}}^2}{2r^2} - {\Lambda}^{(h)}(t,\,\theta,\,\bar \theta)\, (r - a),
\end{eqnarray}
where ${P_r}^{(b)}(t,\,\theta,\,\bar \theta)$ and ${\Lambda}^{(h)} (t, \theta, \bar\theta)$ 
are from (17) and (11). It is straightforward to note that the (anti-)BRST 
invariance of the action integral corresponding to this part of  the Lagrangian can be captured in the following expressions:
\begin{eqnarray}
\frac{\partial}{\partial\theta} \bigl[\tilde L_0\bigr]|_{\bar\theta = 0} &=& -\,\frac{d}{dt}\bigl[\bar C\,(r - a)\bigr]\,
\Longleftrightarrow\, s_{ab}\,L_0  = -\,\frac{d}{dt}\bigl[\bar C\,(r - a)\bigr],\nonumber\\
\frac{\partial}{\partial{\bar\theta}} \bigl[\tilde L_0\bigr]|_{\theta = 0} &=& -\,\frac{d}{dt}\bigl[ C\,(r - a)\bigr]\,
\Longleftrightarrow \,s_{b}\,L_0  = -\,\frac{d}{dt}\bigl[ C\,(r - a)\bigr],\nonumber\\
\frac{\partial}{\partial \bar \theta} \frac{\partial}{\partial{\theta}} \bigl [\tilde L_0 \bigr] &=& - i\,
\frac{d}{dt}\bigl[b\,(r - a)\bigr]\,
\Longleftrightarrow \,s_b s_{ab}\,L_0  = -\,i\, \frac{d}{dt}\bigl[b\,(r - a)\bigr],
\end{eqnarray}
where inputs from the expansions (11) and (17) have been taken into account. Furthermore,
we note that the following supervariable generalizations are trivial:
\begin{eqnarray}
r(t) \longrightarrow R(t, \theta, \bar{\theta}) = r(t), \quad \vartheta(t) \longrightarrow \Theta(t, \theta, \bar{\theta}) = \vartheta(t),
\quad  p_{\vartheta}(t) \longrightarrow P_{\vartheta}(t, \theta, \bar{\theta}) =  p_{\vartheta}(t),
\end{eqnarray}
because of the fact that these variables do {\it not} transform under the (anti-)BRST symmetry
transformations (i.e. $s_{(a)b}\, [r,\, \vartheta,\, p_{\vartheta}] = 0 $). In other words, there is no Grassmannian expansions for 
these variables when they are generalized onto (1, 2)-dimensional supermanifold.

In the above expressions (cf. (38), (39)), we have captured the (anti-)BRST invariance of the starting Lagrangian $L_0$ for 
the rigid rotor in the language of supervariable approach. The gauge-fixing and Faddeev-Popov ghost terms of the 
starting Lagrangian (1):
\begin{eqnarray}
L_{gf} + L_{FP} = b\,(\dot\lambda - p_r) + \frac{b^2}{2} - i\,\dot{\bar C}\,\dot C + i\,
\bar C\,C,
\end{eqnarray} 
can be generalized onto the (1, 2)-dimensional supermanifold as:
\begin{eqnarray}
{\tilde L}_{gf} + {\tilde L}_{FP} = b(t)\,\bigl[{\dot\Lambda}^{(h)} - P^{(b)}_r \bigr] 
+  \frac{b^2 (t)}{2} - i\,\dot{\bar F}^{(h)}\,{\dot F}^{(h)} 
+ i\,{\bar F}^{(h)}\,F^{(h)},
\end{eqnarray} 
where we have taken $b(t) \longrightarrow  B(t, \theta, \bar{\theta})= b (t)$ and the other expansions are given in (11) and (17).
It is straightforward to check that:
\begin{eqnarray}
\frac{\partial}{\partial\theta} \bigl[{\tilde L}_{gf} + {\tilde L}_{FP} \bigr]|_{\bar\theta = 0} &=& \frac{d}{dt}\bigl[b\, \dot{\bar C}\bigr]\,
\equiv \, s_{ab}\,\bigl[{L}_{gf} + {L}_{FP} \bigr],\nonumber\\
\frac{\partial}{\partial\bar\theta} \bigl[{\tilde L}_{gf} + {\tilde L}_{FP} \bigr]|_{\theta = 0} &=& \frac{d}{dt}\bigl[b\, \dot{C}\bigr]\,
\equiv \, s_{b}\,\bigl[{L}_{gf} + {L}_{FP} \bigr],\nonumber\\
\frac{\partial}{\partial\theta}\,\frac{\partial}{\partial\bar\theta} \bigl[{\tilde L}_{gf} + {\tilde L}_{FP} \bigr] 
&=& \frac{d}{dt}\bigl[i\,b\, \dot{b}\bigr]\,\equiv \, s_{b}\,s_{ab}\,\bigl[{L}_{gf} + {L}_{FP} \bigr].
\end{eqnarray} 
Hence, the total (anti-)BRST invariant Lagrangian $L_b = L_0 + {L}_{gf} + {L}_{FP} $ can be expressed as the sum of (38)
and (42) in the supervariable approach (as ${\tilde L}_b = {\tilde L}_0 + {\tilde L}_{gf} + {\tilde L}_{FP} $).
Now, it is straightforward to check that the following are true, namely; 
\begin{eqnarray}
\frac{\partial}{\partial\bar\theta} \bigl[{\tilde L}_{b}\bigr]|_{\theta = 0} &=& \frac{d}{dt}\bigl[b\, \dot{C} - C\,(r -a)\bigr]\,
\equiv \, s_{b}\,\bigl[{L}_{0} \bigr],\nonumber\\
\frac{\partial}{\partial\theta} \bigl[{\tilde L}_{b}\bigr]|_{\bar\theta = 0} &=& \frac{d}{dt}\bigl[b\, \dot{\bar C} - \bar C\,(r -a)\bigr]\,
\equiv \, s_{ab}\,\bigl[{L}_{0} \bigr],\nonumber\\
\frac{\partial}{\partial\bar\theta}\,\frac{\partial}{\partial\theta} \bigl[{\tilde L}_{b}\bigr] &=& \frac{d}{dt}\bigl[i\,b\,\bigl\{\dot{b} - (r -a)\bigr\} \bigr]\,
\equiv \, s_{b}\,s_{ab}\,\bigl[{L}_{0} \bigr].
\end{eqnarray}
Thus, we have captured the  (anti-)BRST invariance of the action $S = \int dt\, L_b$ in the language 
of the supervariables (11) and (17) (obtained after various appropriate restrictions) and Grassmannian derivatives.

Taking the help of mappings in (19) and expansions in (11) and (17), it is straightforward to express 
Lagrangian (36) in the language of supervariable on the (1, 2)-dimensional supermanifold, namely;
\begin{eqnarray}
L_b \longrightarrow {\tilde L_b} &\equiv &  {\tilde L}_{0} + \frac{\partial}{\partial\bar\theta}\,\Bigl[ -\,i\, {\bar F}^{(h)}
\bigl\{({\dot \Lambda}^{(h)} - {P_r}^{(b)}) + \frac{b(t)}{2} \bigr\}\Bigr ]|_{\theta = 0},\nonumber\\
&\equiv &  {\tilde L}_{0} + \frac{\partial}{\partial\theta}\,\Bigl[ i\, F^{(h)}
\bigl\{({\dot\Lambda}^{(h)} - {P_r}^{(b)}) + \frac{b(t)}{2} \bigr\}\Bigr]|_{\bar \theta = 0},\nonumber\\
&\equiv &  {\tilde L}_{0} + \frac{\partial}{\partial\bar\theta}\,\frac{\partial}{\partial\theta}\,
\Bigl[ \frac {i}{2}\,\bigl({\Lambda}^{(h)}\,{\Lambda}^{(h)} - {P_r}^{(b)}\,{P_r}^{(b)} \bigr) + \frac{F^{(h)}\,{\bar F}^{(h)}}{2}\Bigr].
\end{eqnarray}
Using the nilpotency and anticommutativity properties of the translational generators ($\partial_{\theta},\,\partial_{\bar\theta}$), 
it is clear that the (anti-)BRST invariance of the action integral corresponding to the Lagrangian $L_b$ 
can be captured in the language of supervariable approach because 
$(\partial_\theta \, {\tilde L}_b), \,(\partial_{\bar\theta} \, {\tilde L}_b)$ and 
($\partial_{\bar\theta}\,\partial_\theta \,{\tilde L}_b$)
are {\it all} total time derivatives.

We concentrate now on the (anti-)co-BRST invariance of the Lagrangian (1) in the language of the supervariable approach. Here, 
we shall {\it not} be as much elaborate as we have been in the case of (anti-)BRST invariance of the Lagrangian within the framework of supervariable approach. 
We can generalize the Lagrangian (1) to the (1, 2)-dimensional supermanifold in a straightforward manner as:
\begin{eqnarray}
L_b \longrightarrow {\tilde L_b}^{(d)} &=& \dot r\, {P_r}^{(d)} + \dot\vartheta\, p_{\vartheta} - \frac{{ p_{\vartheta}}^2}{2\,r^2} 
- \Lambda^{(d)} (r - a) + b\,({\dot\Lambda}^{(d)} - {P_r}^{(d)})\nonumber\\
 &+& \frac{b^2}{2} - i\, \dot{\bar F}^{(d)}{\dot F}^{(d)} + i\,{\bar F}^{(d)}\,{F}^{(d)},
\end{eqnarray}
where (${\Lambda}^{(d)},\, {P_r}^{(d)},\, {F}^{(d)},\,{\bar F}^{(d)}$) are the expansions (34) that have been 
derived by exploiting the DHC (cf. (21))  and (anti-)co-BRST invariant restrictions. The (anti-)co-BRST invariance of the
starting Lagrangian (1) can be captured within the framework of supervariable approach, in the following fashion:
\begin{eqnarray}
\frac{\partial}{\partial\bar\theta} \bigl[{{\tilde L}_{b}}^{(d)}\bigr]|_{\theta = 0} &=& 0 \,\Longleftrightarrow\, 
s_{d}\,\bigl[{L}_{b} \bigr] = 0,\nonumber\\
\frac{\partial}{\partial\theta} \bigl[{{\tilde L}_{b}}^{(d)}\bigr]|_{\bar\theta = 0} &=& 0 \,\Longleftrightarrow\, 
s_{ad}\,\bigl[{L}_{b} \bigr] = 0,\nonumber\\
\frac{\partial}{\partial\bar\theta}\,\frac{\partial}{\partial\theta} \bigl[{{\tilde L}_{b}}^{(d)}\bigr] &=& 0 \,\Longleftrightarrow\, 
s_d\,s_{ad}\,\bigl[{L}_{b} \bigr] = 0.
\end{eqnarray}
Geometrically, the (anti-)co-BRST invariance can be explained as follows. The super-Lagrangian 
${\tilde L_b}^{(d)}$ is the sum  of composite (super)variables (obtained after DHC and 
appropriate set of (anti-)co-BRST invariant restrictions) such
that its translation along $\theta$ and $\bar\theta$-directions yields zero result (which is equivalent to $s_{(a)d} L_0 = 0$).

\noindent
\section{Nilpotency and anticommutativity: Supervariable approach to a 1D rigid rotor}

In this section, we discuss  the nilpotency and absolute anticommutativity properties of the (anti-)co-BRST 
charges within the framework of supervariable approach. We also {\it briefly} mention about the same properties that are
associated with the nilpotent (anti-)BRST charges because this has been already discussed, 
to some extent, in our earlier work [4]. 
In fact, we shall pinpoint only a few subtle points connected with the (anti-)BRST charges which have {\it not} been 
mentioned in our earlier work [4]. For instance, we shall touch upon the absolute anticommutativity of the BRST and anti-BRST
charges and its geometrical meaning in the language of the translational generators, 
($\partial_\theta,\,\partial_{\bar\theta}$) within the framework of our approach.

To begin with, first of all, we note that the (anti-)co-BRST charges ($Q_{(a)d}$) can be expressed in the following forms 
within the framework of the supervariable approach, namely;
\begin{eqnarray}
Q_{ad} &=& -\,i\,\frac{\partial}{\partial\bar\theta}\,\frac{\partial}{\partial\theta}\,\bigl[ {\dot\Lambda}^{(d)}\, F^{(d)}\bigr]
\qquad\,\, \qquad \quad \equiv  -\,i\, \int d\bar\theta \,\int d\theta \,\bigl[ {\dot\Lambda}^{(d)}\, F^{(d)}\bigr],\nonumber\\
Q_{ad} &=& -\,i\,\frac{\partial}{\partial\bar\theta}\, \bigl [{\dot F}^{(d)} \,F^{(d)}\bigr]|_{\theta = 0}
\quad \, \quad\qquad\quad \equiv  -\,i\,  \int d\bar\theta \, \bigl [{\dot F}^{(d)} \,F^{(d)}\bigr]|_{\theta = 0},\nonumber\\
Q_{ad} &=& i\, \frac{\partial}{\partial\theta}\,\bigl[ \dot{\bar F}^{(d)}\,F^{(d)}
 - i\,{\dot\Lambda}^{(d)} \, \dot b (t) \bigr]|_{\bar\theta = 0}
\quad \equiv\,  i\, \int d\theta\, \bigl[ \dot{\bar F}^{(d)}\,F^{(d)} -
 i\,{\dot\Lambda}^{(d)} \, \dot b (t) \bigr]|_{\bar\theta = 0},\nonumber\\
Q_{d} &=& -\,i\,\frac{\partial}{\partial\bar\theta}\,\frac{\partial}{\partial\theta}\,
\bigl[ {\dot\Lambda}^{(d)}\, {\bar F}^{(d)}\bigr]
\qquad\,\,\, \qquad \quad \equiv  -\,i\, \int d\bar\theta \,\int d\theta 
\,\bigl[ {\dot\Lambda}^{(d)}\, {\bar F}^{(d)}\bigr],\nonumber\\
Q_{d} &=& i\,\frac{\partial}{\partial\bar\theta}\,\bigl [\dot{\bar F}^{(d)} 
\,{\bar F}^{(d)}\bigr]|_{\theta = 0} 
\qquad\,\,\,\,\,\,\,\qquad\quad \equiv  i\,  \int d\bar\theta \, 
\bigl [\dot{\bar F}^{(d)} \,{\bar F}^{(d)}\bigr]|_{\theta = 0},\nonumber\\
Q_{d} &=& -\,i\, \frac{\partial}{\partial\theta}\,\bigl[ \dot{F}^{(d)}\,{\bar F}^{(d)} +
 i\,{\dot\Lambda}^{(d)} \, \dot b (t) \bigr]|_{\bar\theta = 0} \nonumber\\
 &\equiv& \,-\,  i\, \int d\theta \,\bigl[ \dot{F}^{(d)}\,{\bar F}^{(d)} +
 i\,{\dot\Lambda}^{(d)} \, \dot b (t) \bigr]|_{\bar\theta = 0},
\end{eqnarray}
where the super-expansions (34) have been taken into account that have been derived after the application 
of the DHC (cf. (21)) and several other (anti-)co-BRST invariant restrictions. Furthermore, consistent with the 
super-expansions (34), the (anti-)co-BRST charges in (5) have been re-expressed as follows
\begin{eqnarray}
Q_d = \dot r\, \bar C - (r - a)\,\dot{\bar C}, \qquad  \qquad Q_{ad} =  \dot r\, C - (r - a)\,\dot{C},
\end{eqnarray}
where we have used the Euler-Lagrange equations of motion $b = \dot r$ and $\dot b = -\, (r - a)$ 
that emerge from the starting Lagrangian (1) because of the least action principle.
There are some alternative expressions for the ones quoted in (48). For instance, one can replace $ \dot{\Lambda}^{(d)} $ by 
$ P^{(d)}_{r} $ and, once again,  we obtain the same expressions for the (anti-)co-BRST charges
$ Q_{(a)d} $.

Due to the mappings listed in (35), we can express the above expressions (48) in the ordinary space in the
language of the   nilpotent and absolutely anticommuting (anti-)co-BRST transformations $ s_{(a)d} $ and ordinary variables as:
\begin{eqnarray}
&& Q_{ad} = -\,i\,s_d\, s_{ad}\, \bigl[ {\dot\lambda}\, C\bigr], \qquad\qquad\quad Q_{d} =
 -\,i\,s_d\, s_{ad}\, \bigl[ {\dot\lambda}\, \bar C \bigr], \nonumber\\
&& Q_{ad} = -\,i\,s_d\, \bigl[\dot C \, C \bigr], \qquad\qquad \qquad \, Q_{d} =
 \,i\,s_{ad}\, \bigl[\dot {\bar C} \, {\bar C} \bigr],\nonumber\\
&& Q_{ad} = i\, s_{ad}\, \bigl[\dot{\bar C}\,C - i\, \dot\lambda\, \dot b\bigr] \,\qquad\qquad 
Q_{d} = -\,i\,s_d\,\bigl[\dot C\, \bar C + i\, \dot\lambda\,\dot b\bigr]. 
\end{eqnarray}
By exploiting the (anti-)co-BRST symmetry transformation (4), it can be checked that the above expressions 
do match with (49) (which is also equivalent to expressions given in (5) in terms of the auxiliary variable $b(t)$).
From the above equations, it becomes transparent that the nilpotency of (anti-)co-BRST charges is deeply connected
with the nilpotency ($s_{(a)d}^2  = 0$) of (anti-)co-BRST symmetry transformations as well as the nilpotency 
(${\partial_\theta}^2 = 0,\, {\partial_{\bar\theta}}^2 = 0 $) of the 
translational generators $\partial_\theta$ and $\partial_{\bar\theta}$ along 
the Grassmannian directions of this (1, 2)-dimensional
supermanifold. For instance, if we consider 
$Q_{d} = -\,i\,s_d\,\bigl[\dot C\, \bar C + i\, \dot\lambda\,\dot b\bigr]$, 
it is clear that $s_d \, Q_d = i\, \{Q_d,\,Q_d \} = 0$ because of $s_{d}^2 = 0$ and the basic definition of a 
generator of a given transformation. Furthermore, from the suitable
expressions from (48), it is very evident that $\partial_{\bar\theta}\, Q_d = 0$ due to $ {\partial_{\bar\theta}}^2 = 0$ 
which, in turn, implies that ${Q_d}^2 = 0$. Such kind of arguments can be also given for the nilpotency of 
$Q_{ad}$ as well. Geometrically, the equation 
$ Q_{d} = -\,i\, \frac{\partial}{\partial\bar\theta}\,\bigl[ \dot{F}^{(d)}\,{\bar F}^{(d)} 
+ i\,{\dot\Lambda}^{(d)} \, \dot b (t) \bigr]|_{\theta = 0}$
implies that the co-BRST charge $ Q_d $ is already equivalent to the translation of a composite supervariable 
$ (\dot{\bar F}^{(d)} \,{\bar F}^{(d)}) $ along the $ \bar{\theta} $-direction of the supermanifold. Thus, any further 
translation along $ \bar{\theta}$-direction produces a zero result because of the fermionic 
(${\partial_{ \bar{\theta}}}^{2} = 0 $) nature of $ \partial_{\bar{\theta}} $. Similar explanation for the 
nilpotency of the suitable expression for $Q_{ad}$ can be given in the language of nilpotency 
(${\partial_{\theta}}^{2} = 0 $) of the translational generator $\partial_{\theta}$ along $\theta$-direction.

Now we dwell a bit on the geometrical meaning of the absolute anticommutativity of the (anti-)co-BRST charges $ Q_{ad} $
 in the language of the translational generators ($\partial_{\theta} $ and  $\partial_{ \bar{\theta}}$) 
along the Grassmannian directions of the supermanifold. Let us take the first example as:
\begin{eqnarray}
Q_{d} &=& \,i\, \frac{\partial}{\partial\theta}\,\bigl[ \dot{\bar F}^{(d)}\,{\bar F}^{(d)} 
 \bigr]|_{\bar\theta = 0} 
\equiv  \, i \, s_{ad}\,\bigl[\dot{\bar C}\,\bar{C}\bigr].
\end{eqnarray}
It is self-evident that $ s_{ad}\,Q_{d} = 0 $ because of the nilpotency ($s_{ad}^2 = 0$) of $ s_{ad} $ and 
$ \partial_{\theta}\,Q_{d}  =0 $ because of the nilpotency $ ( {\partial_{\theta} }^{2} = 0 ) $ of the translational 
generator  $\partial_{\theta}$. However, if we take the definition of the generator for the transformation $ s_{ad} $,
 then, $ s_{ad}\,Q_{d} = i\, \left\lbrace Q_{d}, Q_{ad} \right\rbrace = 0$ due to the nilpotency ($ s_{ad}^{2} = 0 $)
of $ s_{ad} $ which in turn implies the absolute anticommutativity of the (anti-)co-BRST charges $ Q_{(a)d} $. 
 If we operate by a $ \partial_{\bar\theta} $ on (51),
we should get $ \partial_{\bar{\theta}}\,Q_d = 0 $. However, it leads to the following explicit expressions:
\begin{eqnarray}
\frac{\partial}{\partial\bar{\theta}}\, Q_{d} &=& 0 \,=\, \,i\,\frac{\partial}{\partial\bar{\theta}}\,
\frac{\partial}{\partial\theta}\,\left[ \dot{\bar F}^{(d)}\,\bar{F}^{(d)}\right]
\nonumber\\
& \equiv & \,\frac{i}{2}\,\left( \partial_{\theta}\,\partial_{\bar\theta} + \partial_{\bar{\theta}}\,\partial_{\theta} \right)\,
\left[\dot{\bar F}^{(d)}\,\bar{F}^{(d)}\right],    
\end{eqnarray}
which shows the absolute anticommutativity of the (anti-)co-BRST charges because of the fact that 
$ \partial_{\theta}\,\partial_{\bar\theta} + \partial_{\bar{\theta}}\,\partial_{\theta}  = 0$.
If we take into account the mappings listed in (35), we obtain 
$ s_{d}\,s_{ad} + s_{ad}\,s_{d} = 0 $. The latter is equivalent to the absolute anticommutativity of  the
(anti-)co-BRST charges. On the other hand, from (51), it is clear that $ \partial_{\theta}\,Q_{d} =  0$ because
the nilpotency of $ \partial_{\theta}$ (i.e. $ \partial^{2}_{\theta} = 0$). Thus, we observe that the nilpotency
and anticommutativity properties are inter-related. These observations are true because the nilpotency condition 
$ (\partial^{2}_{\theta} =  \partial^{2}_{\bar\theta} = 0 $) is a limiting case of the absolute anticommutativity 
($\partial_{\theta}\,\partial_{\bar\theta} + \partial_{\bar{\theta}}\,\partial_{\theta}  = 0$).
This is due to the fact that (i) when we take  $\partial_{\theta} = \,\partial_{\bar\theta}$, 
we obtain  $ \partial^{2}_{\bar\theta} = 0$, and (ii) when we choose  $\partial_{\bar\theta} =\partial_{\theta} $, 
we get $ \partial^{2}_{\theta} = 0$. Similar inferences could be drawn for the nilpotency 
$ (s^{2}_{a(d)} = 0)$ of the (anti-)co-BRST symmetry transformations ($ s_{(a)d} $) 
(and their corresponding charges $ Q_{(a)d}$ ) 
from the absolute anticommutativity $ s_{d}\,s_{ad} + s_{ad}\,s_{d} = 0 $ 
(and their counterparts $ Q_{d}\,Q_{ad} + Q_{ad}\,Q_{d} = 0 $).

We close this section with a brief remark about the absolute anticommutativity 
($ s_{b}\,s_{ab} + s_{ab}\,s_{b} = 0, \, Q_{b}\,Q_{ab} + Q_{ab}\,Q_{b} = 0  $) of the (anti-)BRST
symmetries (and their corresponding charges $ Q_{(a)b} $) which have been discussed in our earlier work [4]
within the framework of supervariable approach. For instance, we have obtained the results 
$ Q_{b} = i\, s_{ab}\,(C\,\dot{C}),\, Q_{ab} = -\,i\, s_{b}\,(\bar{C}\,\dot{\bar{C}}) $ and their corresponding
expressions in the supervariable approach. Now, it is crystal clear that 
$s_{ab}\,Q_{b} = i\, \left\lbrace Q_{b}, \, Q_{ab}\right\rbrace  = 0$
due to the nilpotency $(s^{2}_{ab} = 0)$ of $ s_{ab} $. Similarly, 
$ s_{b}\,Q_{ab} = i\, \left\lbrace Q_{ab}, \, Q_{b}\right\rbrace  = 0 $ due to the 
nilpotency $(s^{2}_{b} = 0)$ of the BRST symmetry transformations $ s_{b} $. Thus, we note that 
the absolute anticommutativity of  (anti-)BRST charges is connected with the {\it nilpotency} $(s^{2}_{(a)b} = 0)$ of 
the (anti-)BRST symmetry transformations $s_{(a)b}$. These observations are {\it logical} because, 
as discussed earlier, the absolute anticommutativity 
($  \partial_{\theta}\,\partial_{\bar\theta} + \partial_{\bar{\theta}}\,\partial_{\theta}  = 0$) 
of the translational generators ($  \partial_{\theta}, \,\partial_{\bar\theta} $) 
is connected with the nilpotency ($ \partial^{2}_{\theta} = 0 = \partial^{2}_{\bar\theta} $) 
of these translational operators is the limiting cases when $ \partial_{\theta} =  \partial_{\bar\theta} $ 
and/or $  \partial_{\bar\theta} =  \partial_{\theta} $.

\section{Conclusions}

In our present endeavor, we have derived the (anti-)BRST symmetry transformations by exploiting 
the ideas of (i) horizontality condition, and (ii) (anti-)BRST invariant restrictions, on the supervariables
which are defined on the suitably chosen (1, 2)-dimensional supermanifold (on which our ordinary
theory is generalized). These ideas are geometrically and physically more intuitive as well as elegant
and the {\it latter} condition is completely different from our earlier work [4] where mathematically
correct (but ad-hoc) approximations have been made. In our present investigation, the geometrical
interpretation for the nilpotency and anticommutativity properties, associated with the (anti-)BRST
charges, remain the same as has been discussed in our earlier work  [4] on this topic.

One of the relatively novel features of our present investigation is the systematic application
of the DHC for the precise derivation of the proper (anti-)co-BRST symmetry transformations where the 
Hodge duality ($\star$) operation on the (1, 2)-dimensional supermanifold plays a very decisive role.
We have verified that the working-rules, laid down in [14], turn out to be
correct because we are able to derive the precise form of the 
nilpotent (anti-)co-BRST symmetry transformations 
in a consistent manner.  We have also provided the geometrical basis for the 
(anti-)co-BRST charges in the language of the supervariables (obtained  after the application
of the appropriate set of restrictions) and the translational generators along the Grassmannian
directions of the supermanifold.

It is very important for us to apply the key ideas of DHC (and associated Hodge duality $\star$
operation) in the context of the other higher dimensional physical systems of interest (that have been 
proven to be the tractable physical examples of Hodge theory) so that the working-rules, laid down
in [14], could be tested on any arbitrary (D, 2)-dimensional  supermanifold. For instance, we have
already discussed the utility of the Hodge duality $\star$ operation on the (4, 2)-dimensional
supermanifold in the case of 4D Abelian gauge theory in our earlier work [14]. Thus, the application
of the DHC (in the context of some physical systems of interest) remains a central issue for our future 
endeavors. It is gratifying to state that we  have already applied the DHC in the cases of 
the modified versions of 2D Proca theory as well as the chiral bosonic field theory and 
have obtained the precise form of the (anti-)co-BRST symmetries [15, 16].
We are currently busy with the ideas of the application of DHC and our results would be reported
in our future publications. \\

\noindent
{\bf Conflict of Interests} \\

\noindent
The authors declare that there is no conflict of interests
as far as the publication of this paper is concerned.

\noindent
{\bf Acknowledgements} \\

\vskip 0.5cm
\noindent
TB is grateful to the BHU-fellowship and DS thanks UGC,
Government of India, New Delhi, for the financial support through RFSMS scheme under which 
the present investigation has been carried out. Fruitful discussions with our esteemed Reviewer,
on the topic of the subject matter of our paper, are thankfully acknowledged, too.\\

\vskip 1cm

\begin{center}
\Large{\bf Appendix}\\
\end{center}
\vskip 1cm
We compute here the explicit expression for $\star\,\tilde d\,\star\,{\tilde\lambda}^{(1)}$
 which has been used in the DHC (21). Towards this goal in mind, we exploit the working-rule,
developed in [14], for the Hodge duality operation on a (1, 2)-dimensional supermanifold.
To begin with, we have the following single ($\star$) operation on the super 1-form:
\begin{eqnarray}
\star\,{\tilde\lambda}^{(1)} = \star\,(dt\,\Lambda + d\theta\,F + d\bar\theta\, F).
\end{eqnarray}
According to the working-rule laid down in [14], we have the following correct $(\star)$ operation on
the 1-form differentials of the (1, 2)-dimensional supermanifold, namely;
\begin{eqnarray}
\star\,(dt) = (d\theta \wedge d\bar\theta), \qquad \star\,(d\theta) = (dt \wedge d\bar\theta),
\qquad \star\,(d\bar\theta) = (dt \wedge d\theta).
\end{eqnarray}
The above expressions {\it physically} imply that, on the (1, 2)-dimensional supermanifold, the {\it dual}
of the differential ($dt$) is ($d\theta \wedge d\bar\theta$). In exactly similar fashion, 
the {\it dual} of the differentials ($d\theta$) and ($d\bar\theta$) have been expressed 
(taking into account the physical arguments). These inputs imply the following expression
for the super 2-form that is derived from  (53):
\begin{eqnarray}
\star\,{\tilde\lambda}^{(1)} = (d\theta \wedge d\bar\theta)\,\Lambda + 
(dt \wedge d\bar\theta)\,F + (dt \wedge d\theta)\,F .
\end{eqnarray}
Now, we have to operate $\tilde d = dt\,\partial_t + d\theta  \,\partial_{\theta} +
 d{\bar\theta}\,\partial_{\bar\theta}$ on it. As a consequence of this operation, we obtain the
following super 3-form:
\begin{eqnarray}
\tilde d\,\star\,{\tilde\lambda}^{(1)} &=& (dt \wedge d\theta \wedge d\bar\theta)\,{\dot\Lambda} 
+ (dt \wedge d\bar\theta \wedge dt)\,\dot{\bar F} + 
(dt \wedge dt \wedge d\theta)\,\dot F\nonumber\\ &+& 
(d\theta \wedge d\theta \wedge d\bar\theta)\,\partial_{\theta}\,\lambda 
- (d\theta \wedge dt \wedge d\bar\theta)\,\partial_{\theta}\,\bar F - 
(d\theta \wedge dt \wedge d\theta)\,\partial_{\theta}\, F\nonumber\\
&+& (d\bar\theta \wedge d\theta \wedge d\bar\theta)\,\partial_{\bar\theta}\,\lambda 
- (d\bar\theta \wedge dt \wedge d\bar\theta)\,\partial_{\bar\theta}\,\bar F
- (d\bar\theta \wedge dt \wedge d\theta)\,\partial_{\bar\theta}\, F.
\end{eqnarray}
To fully calculate $\star\,\tilde d\,\star\,{\tilde\lambda}^{(1)}$, we have to operate another 
($\star$) on the above super 3-form to obtain a super 0-form. Before we carry out the above 
operation, it is clear that the {\it second} and {\it third} terms of the top line in (56) 
would be zero due to $(dt \wedge dt = 0)$. Further, as the working-rules laid down in [14], 
the 3-forms with {\it only}  Grassmannian differentials would be 
zero on the (1, 2)-dimensional supermanifold because {\it physically} such a supermanifold cannot 
accommodate a super 3-form that is expressed in terms of the wedge products of {\it three}
Grassmannian variables {\it only}. 
Physically, the allowed super 3-form differential wedge products on the (1, 2)-dimensional
supermanifold are: $(dt \wedge d\theta \wedge d\bar\theta), (dt \wedge d\theta \wedge d\theta),
(dt \wedge d\bar\theta \wedge d\bar\theta)$ because these contain the wedge products that incorporate
one differential ($dt$) of bosonic nature and two differentials [i.e. ($d\theta \wedge d\theta$), 
($d\theta \wedge d\bar \theta$) and ($d\bar\theta \wedge d\bar\theta$)] of the fermionic nature. 
These arguments imply that the {\it fourth} and {\it seventh} terms would be zero. 
To be more precise, we note that the coefficients of 3-form differential wedge-products 
($d\theta \wedge d\theta \wedge d\bar\theta$) and ($d\bar\theta \wedge d\theta \wedge d\bar\theta$) 
do {\it not} contribute to the derivation of the proper (anti-) co-BRST symmetries.
Thus, these terms are {\it not} physically important.
As a consequence, only the following terms would, ultimately, exist in (56), namely;
\begin{eqnarray}
\tilde d\,\star\,{\tilde\lambda}^{(1)} &=& (dt \wedge d\theta \wedge d\bar\theta)\,{\dot\Lambda}
+ (dt \wedge d\theta \wedge d\bar\theta)\,\partial_{\theta}\,\bar F +
 (dt \wedge d\theta \wedge d\theta)\,\partial_{\theta}\, F\nonumber\\
&+& (dt \wedge d\bar\theta \wedge d\bar\theta)\,\partial_{\bar\theta}\,\bar F
+(dt \wedge d\theta \wedge d\bar\theta)\,\partial_{\bar\theta}\, F.
\end{eqnarray}
It is worth pointing out that, mathematically, any arbitrary number of differentials may exist
 in the wedge product with {\it only} the Grassmannian differentials 
(e.g. $d\theta  \wedge d\theta \wedge d\theta \wedge d\theta,\;
d \bar \theta  \wedge d\bar \theta \wedge d\bar \theta \wedge d\bar\theta ... $) etc. 
However, physically, it is not permitted to have any arbitrary number of wedge
products of the Grassmannian differentials on a given {\it finite} ($D$, 2)-dimensional 
supermanifold
 on which a $D$-dimensional ordinary physical theory is generalized. Thus, the
derivation of (57) is physically correct. Now, the stage is set to 
apply another ($\star$)  on it. Using the following inputs 
(see, e.g. [14]) on the (1, 2)-dimensional supermanifold
\begin{eqnarray}
\star\,(dt \wedge d\theta \wedge d\bar\theta) = 1,\qquad \star\,(dt \wedge d\theta \wedge d\theta)
 = s^{\theta\,\theta},
\qquad \star\,(dt \wedge d\bar\theta \wedge d\bar\theta) = s^{\bar\theta\,\bar\theta},
\end{eqnarray}
where $s^{\theta\,\theta}$ and $ s^{\bar\theta\,\bar\theta}$ are symmetric in $\theta$ and $\bar\theta$ indices, 
we obtain the final expression
\begin{eqnarray}
\star\,\tilde d\,\star\,{\tilde\lambda}^{(1)} = (\dot\lambda + \partial_\theta \,\bar F + \partial_{\bar\theta}\,F)
 +  s^{\bar\theta\,\bar\theta}\, \partial_{\bar\theta}\,\bar F + s^{\theta\,\theta}\,\partial_{\theta}\,F,
\end{eqnarray}
which is used in the main body of our text (cf. (22)). The first entry of the equation (58) {\it physically} 
implies that the dual of the wedge product $(dt \wedge d\theta \wedge d\bar\theta)$ is nothing but 
{\it unity} (i.e. a $0$-form) as all the {\it three} independent differentials of the (1, 2)-dimensional
supermanifold are present in it. On the other hand, the dual of $(dt \wedge d\theta \wedge d\theta)$
has been taken to be $s^{\theta\,\theta}$ (i.e. a $0$-form) because when we take another ($\star$) 
operation on it, we should get back the original wedge product $(dt \wedge d\theta \wedge d\theta)$ 
 modulo a sign factor. Similar is the argument for the definition of the duality operation on 
the super 3-form wedge product  $(dt \wedge d\bar\theta \wedge d\bar\theta)$.

We would like to end this Appendix with the remarks that another Hodge duality ($\star$)
operation on (54) is as follows:
\begin{eqnarray}
&&\star \,\, [\,\star\, \,(dt)] = \star \, (d\theta \wedge d\bar\theta) = dt, \qquad
\star \, \,[\,\star\,\, (d\theta)] = \star \, (dt \wedge d\bar\theta) = d\theta, \nonumber\\
&&\star \, \,[\,\star\,\, (d\bar\theta)] = \star \, (dt \wedge d\theta) = d\bar\theta.
\end{eqnarray}
Physically, a single Hodge duality operation on the super 2-form differentials
($d\theta \wedge d\bar\theta$) would be {\it dual} of this wedge product on a (1, 2)-dimensional
supermanifold. It is self-evident that it should be a 1-form. Since the {\it dual}
direction of ($\theta, \bar\theta$) is $t$ on a (1, 2)-dimensional supermanifold, it is 
clear that the resulting 1-form of the dual of ($d\theta \wedge d\bar\theta$) would
be nothing but $dt$. Similar explanation can be given for the other double ($\star$)
operations on the above 1-form differentials. We would like to lay emphasis on the importance of the
factors $s^{\theta\,\theta}$ and $s^{\bar\theta\,\bar\theta}$ in the duality operation in 
Eqn. (58). Their presence, on the r.h.s., gives the idea that when we shall take another ($\star$)
operation on the super 3-forms (in (58)), we shall get back the 
original super 3-forms (modulo some sign factors), namely;
\begin{eqnarray}
&&\star \,\, [\,\star\, (dt \wedge d\theta \wedge d\theta)] = \star \, s^{\theta\,\theta} = 
(dt \wedge d\theta \wedge d\theta), \nonumber\\
&&\star \,\, [\,\star\, (dt \wedge d\bar\theta \wedge d\bar\theta)] = \star \, s^{\bar\theta\,\bar\theta} = 
(dt \wedge d\bar\theta \wedge d\bar\theta), \nonumber\\
&&\star \,\, [\,\star\, (dt \wedge d\theta \wedge d\bar\theta)] = \star \, [\,1 \,] = 
(dt \wedge d\theta \wedge d\bar\theta).
\end{eqnarray}
It is clear that the presence of $s^{\theta\,\theta}$ and $s^{\bar\theta\,\bar\theta}$
{\it do} help us in getting the original super 3-forms after the application of a couple of successive
Hodge duality operations.  
We have not got any sign factors on the r.h.s. (other than ($+$) sign) because of the
fact that we have discussed the {\it double} duality operations on a (1, 2)-dimensional 
supermanifold. However, we do get ($\pm$) signs, after the above kind of double duality operations,
on the (2, 2)-dimensional supermanifold (see, e.g. [14] for details). In a very recent work [17],
the Hodge duality operation on a supermanifold has been discussed in a very elegant manner 
because of the fact that a whole lot of deep mathematical concepts have been taken into account. 
We are sure that the contents of this work [17] are important and they will be very useful
for us in our future work (when we shall take into account the supermanifolds which
would {\it not} be necessarily flat). For our present endeavor, however, we feel that the material
contained, in our earlier work [14] for the {\it flat} (1, 2)-dimensional supermanifold,
is good enough. \\

\end{document}